# Explainable machine learning to enable high-throughput electrical conductivity optimization and discovery of doped conjugated polymers

Ji Wei Yoon[1*], Adithya Kumar[2], Pawan Kumar[2], Kedar Hippalgaonkar[2,3], J Senthilnath[1], Vijila Chellappan[2,4*]

[1]*Institute for Infocomm Research (I²R), Agency for Science, Technology and Research (A*STAR), 1 Fusionopolis Way, #21-01, Connexis South Tower, Singapore 138632, Republic of Singapore*
[2]*Institute of Materials Research and Engineering, Agency for Science Technology and Research, (A*STAR),* 2 Fusionopolis Way, #08-03, Innovis, Singapore 138634, Republic of Singapore
[3]*Materials Science and Engineering, Nanyang Technological University, 50 Nanyang Avenue, Singapore 639798, Singapore*
[4]*Institute for Functional Intelligent Materials, National University of Singapore, Singapore 117544, Singapore*

*\*Corresponding authors*

**Abstract**

The combination of high-throughput experimentation techniques and machine learning (ML) has recently ushered in a new era of accelerated material discovery, enabling the identification of materials with cutting-edge properties. However, the measurement of certain physical quantities remains challenging to automate. Specifically, meticulous process control, experimentation and laborious measurements are required to achieve optimal electrical conductivity in doped polymer materials. We propose a ML approach, which relies on readily measured absorbance spectra, to accelerate the workflow associated with measuring electrical conductivity. The classification model accurately classifies samples with a conductivity >~25 to 100 S/cm, achieving a maximum of 100% accuracy rate. For the subset of highly conductive samples, we employed a regression model to predict their conductivities, yielding an impressive test $R^2$ value of 0.984. We tested the models with samples of the two highest conductivities (498 and 506 S/cm) and showed that they were able to correctly classify and predict the two extrapolative conductivities at satisfactory levels of errors. The proposed ML-assisted workflow results in an improvement in the efficiency of the conductivity measurements by 89% of the maximum achievable using our experimental techniques. Furthermore, our approach addressed the common challenge of the lack of explainability in ML models by exploiting bespoke mathematical properties of the descriptors and ML model, allowing us to gain corroborated insights into the spectral influences on conductivity. Through this study, we offer an accelerated pathway for optimizing the properties of doped polymer materials while showcasing the valuable insights that can be derived from purposeful utilization of ML in experimental science.

**Keywords**

1. Introduction

The time it takes for inventions derived from academic research to become market-ready products spans decades, typically ranging from 20 to 70 years for low-carbon technologies [1]. However, there is a growing demand to expedite various stages of the innovation pipeline in order to effectively tackle urgent societal challenges such as climate change, healthcare requirements, and the development of materials for technological applications. Traditionally, the processes carried out in academic laboratories have been characterized by a high degree of manual labor. However, recent advancements in technology, such as the introduction of robotic systems and the digitalization of these processes, have ushered in a new era of accelerated experimentation and discovery. This transformation is exemplified by the adoption of high-throughput experimentation platforms [2–8], which have revolutionized the speed at which experiments can be conducted. To fully harness the efficiency gains offered by these automated systems, the integration of intelligent algorithms has become crucial. These algorithms empower researchers to make informed decisions regarding experiment planning and execution, maximizing the potential benefits of this automated approach.

A particular class of material systems which could benefit from such an accelerated approach is conjugated polymers. Conjugated polymers, also known as conductive or $\pi$-conjugated polymers, are a unique class of materials with alternating single and double bonds along their backbone structure. This arrangement of bonds creates an extended $\pi$-electron system, giving conjugated polymers their distinctive electronic and optical properties. These materials exhibit semiconducting behavior, making them highly attractive for a wide range of applications, including electronics, optoelectronics, energy storage, sensors, and biomedical devices. Electrical conductivity of these polymers plays a critical role in determining their performance and functionality in various electronic and optoelectronic devices. For example, higher electrical conductivity translates to faster and more reliable operation in transistor, enabling faster data processing and better device performance. Higher electrical conductivity in the active layer of the solar cell allows for improved charge separation, reduced recombination losses, and increased power conversion efficiency. In polymeric thermoelectrics, electrical conductivity is closely linked to the Seebeck coefficient, which characterizes the ability of a material to generate an electric voltage in response to a temperature gradient [9–13]. To optimize the electrical conductivity in conjugated polymers, various strategies such as molecular engineering, doping, controlling morphology, and tuning charge carrier concentration and mobility were adopted [14,15]. Doping is a common approach where dopant molecules are introduced into the polymer matrix to increase charge carrier concentration. However, achieving controlled doping levels and distribution throughout the polymer film is a complex task that requires careful doping strategies and process optimization. In addition, polymeric materials are susceptible to degradation over time, which can affect their electrical conductivity. Factors such as exposure to moisture, oxygen, or high temperatures can lead to chemical and physical changes, resulting in reduced conductivity and increased measurement uncertainty. Given the complexity and time-sensitive nature of optimizing electrical conductivity of conjugated polymers, they have traditionally been prepared and characterized in a laborious fashion using substantial expert knowledge. Recently, ML approaches have emerged as valuable tools to accelerate the workflow by alleviating the high demand on labor and expertise.

The traditional workflow (**Figure 1** (top branch)) associated with most experimental material discovery starts with the design of experiment, where various experimental parameters are selected typically using experts' domain knowledge. The expert would base their choice of parameters either on previous experience or from literature. Then, a manual process would commence, where a human agent prepares samples and then characterizes them. The results of the characterization, coupled with domain knowledge, would inform the next set of points of the parameter space to be sampled. The loop would continue till the discovery of a better material, or when some other exit criteria are met. In our work, we focus on introducing machine learning (ML) models to reduce the inefficiencies associated with the human agent (**Figure 1** (bottom branch)). We introduce the models with two goals in mind: (1) to filter out samples which are deemed uninteresting for further time-consuming characterization, (2) to supplant the expensive characterization with surrogate ML models.

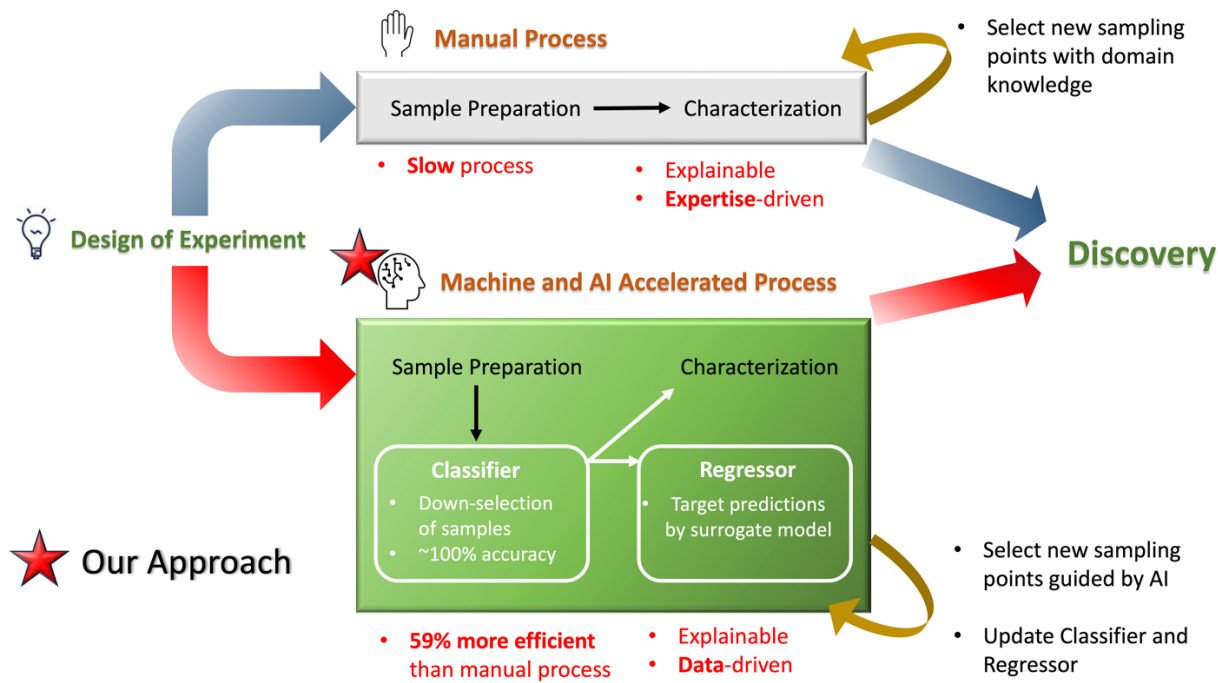

**Figure 1.** Schematic of the traditional versus machine learning-accelerated processes to discovering new materials. Our proposed two-step classification and regression workflow to accelerate laborious measurements is highlighted in the "Machine and AI Accelerated Process" box.

In our work, we aimed to optimize the electrical conductivity of doped conjugated polymers by leveraging on the common electronic physics of materials. Specifically, it is known that an increase in carrier mobility and density of a material would generally lead to a higher conductivity. Also, the higher the carrier density, the larger the absorbance of electromagnetic energy by the carriers, which directly manifest as higher peaks at specific energies in the absorbance spectra. Therefore, we attempt to elucidate the causal relationship between absorbance spectral descriptors and electrical conductivity through the utilization of machine learning techniques. To alleviate the need of an expert agent in the process, our method was designed to be purely data-driven.

For this study, we chose a few high-performing p-type conjugated polymers that were doped with different acceptor molecules. We varied the process parameters, including solvents, doping methods, and dopant concentrations, to create thin films with diverse morphologies and optical/electrical properties. The conjugated polymers used in this investigation are poly (3-

hexylthiophene) (P3HT), poly (2,5-bis(3-tetradecylthiophen-2-yl)thieno[3,2-b]thiophene) (PBTTT-C12 and PBTTT-C14), and quaterthiophene (PDPPP4T). The chemical structure and energy levels of these polymers were extracted from literature [16–19] and are shown in **Figure 2(a)**. These polymers are commonly used as an active layer for devices such as organic solar cells, field-effect transistors, light-emitting diodes, supercapacitors, and sensors [20–28]. We used the common dopant molecules, tris(pentafluorophenyl)borane (BCF), 2,3,5,6-Tetrafluoro-7,7,8,8-tetracyanoquinodimethane (F4TCNQ) and $FeCl_3$, to tune the electrical conductivity. The chemical structure and HOMO/LUMO levels if the polymers as well as the electron affinity of the dopant molecules, extracted from literature are shown in **Figure 2 (a)**. The fabrication method, process variations and measurements used are shown in **Figure 2 (b), (c) and (d)** respectively.

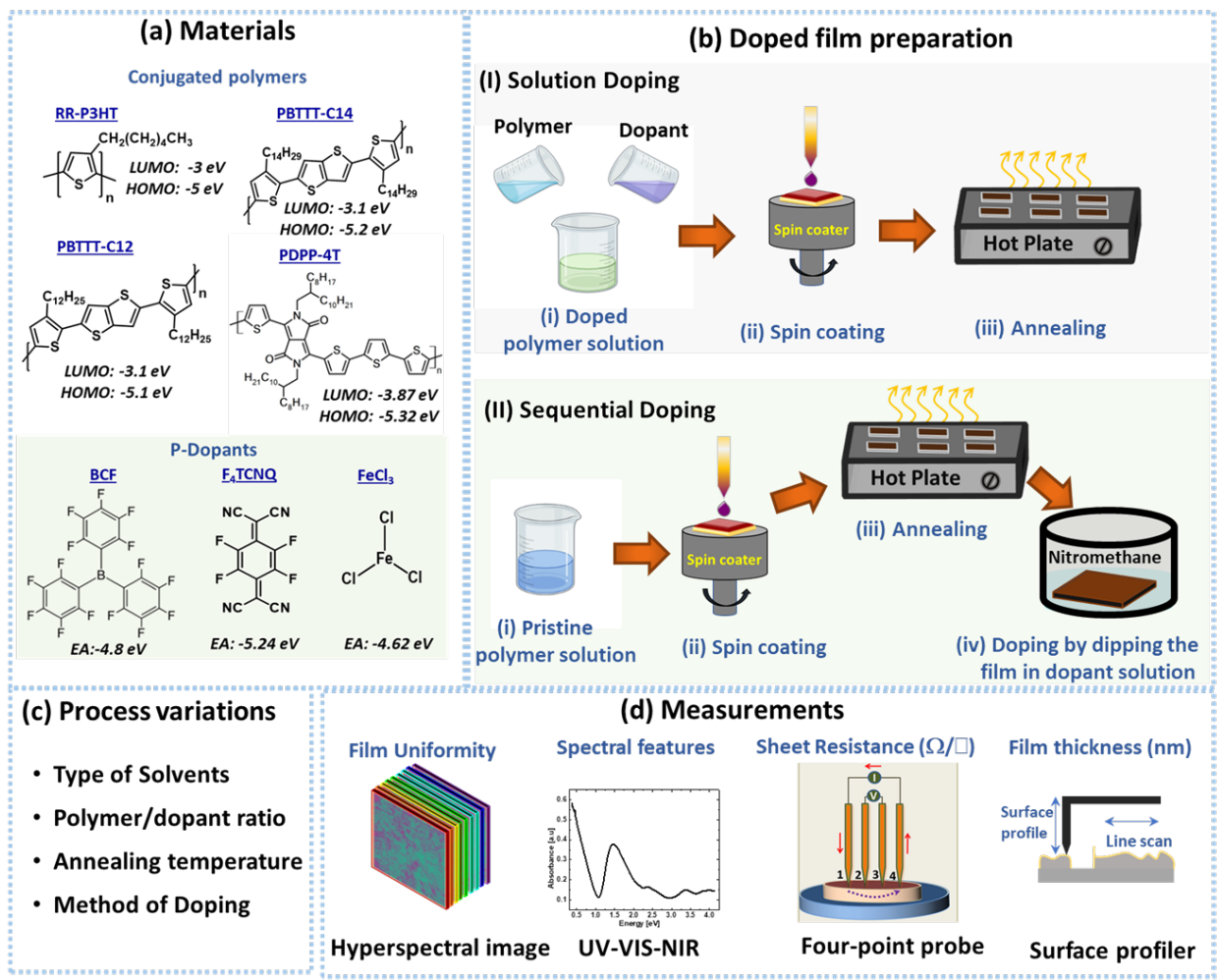

**Figure 2.** Experimental workflow showing the **(a)** details of the materials, **(b)** thin film doping methods, **(c)** process variations and **(d)** measurements used in this study.

The measurement of electrical conductivity typically involves a combination of a four-point probe and a surface profiler. These techniques are used to determine the sheet resistance and film thickness, respectively. However, these methods have their drawbacks as they are destructive to the films and time-consuming, especially when measuring film thickness. Furthermore, relying on a single point of probing may not accurately represent the electrical conductivity of the film, especially if the film is inhomogeneous. Conversely, non-destructive optical techniques offer an alternative solution. By employing spatially resolved spectroscopy methods like hyperspectral imaging, the entire thin film can be probed, providing rapid and comprehensive information about the optical and electrical characteristics of the films [29]. Recently, we demonstrated that the polaronic spectral descriptors and the electrical properties of doped thiophene films can be correlated where the absorption ratio between polarons and neutral excitons exhibited a linear relationship with carrier density [30]. Specifically, Bash et al. [31] conducted optimization of electrical conductivity of P3HT-CNT composites, where they incorporated specific point data from the absorbance spectra in a regression model to predict electrical conductivity. Mamede et al. [32] conducted a forward prediction task in which the UV-VIS absorption of organic molecules was classified using regression algorithms using molecular descriptors and fingerprints. These findings highlight the potential of optical methods for studying electrical conductivity in thin films without causing damage to the samples and enabling quicker data acquisition. In our current work, we take a different approach by using the complete optical spectra (ranging from 300 to 3300 nm) of various doped polymers prepared with different process combinations. This investigation incorporates machine learning techniques, facilitating a deeper understanding of the relationship between optical and electrical properties, and providing more comprehensive insights into thin film behavior.

Across a broader context, ML algorithms have already been successfully applied to various optimization problems. On the experimental front, a self-driving laboratory, which uses the q-expected hypervolume improvement qEHVI algorithm, successfully discovers new synthesis conditions that yield metallic films at lower processing temperatures (below 200 °C) relative to the prior art (250 °C) [8]. In another experimental work [6], a classification model on chemical vapor grown MoS2 is developed, capable of optimizing the synthesis conditions to achieve a higher success rate. Also, a regression model is constructed on the growth of carbon quantum dots, which enhanced the process-related properties, e.g. the photoluminescence quantum yield.

Importantly, ML models have also been applied to computational materials workflows, where the explorable space of materials is typically larger than the experimental. On the computational front, it has been demonstrated that Bayesian optimization using a graph deep learning energy model can be used to perform in-silico relaxations of crystal structures [5]. The authors have been able to predict and synthesize two novel ultra-incompressible hard $MoWC_2$ (P63/mmc) and ReWB (Pca21).

We focus on utilizing ML to minimize human labor in the process. Leveraging the understanding of how UV-VIS-NIR absorbance spectral descriptors (specifically absorption intensity and peak position) change with the introduction of charge carriers or doping, we posit that ML models with good performance can be developed by capturing the correlation between conductivity and the UV-VIS-NIR absorbance spectra. To this end, we will propose a new basis set of local spectral descriptors for such learning tasks based on B-splines. We will show the advantage of using the set of physics-inspired descriptors over other naïve choices, e.g., descriptors that are based solely on global statistical descriptors of the spectra. We note that, despite the lack of multiple factors that influence the conductivity of polymers (refer to [33] for a comprehensive discussion) our trained model still exhibits remarkable performance.

Many ML models are black-box in nature, where their internal computations are typically hidden. The black-box nature of ML is problematic in science because it is challenging to understand how variables interact to make predictions. In ML, black-box models are created directly from data by an algorithm, so that even its designer would not be able to understand how it works. This lack of transparency can lead to problems such as bias and discrimination [34]. In scientific research, such models would only serve as a prediction tool but would not provide insights that are crucial for scientific understanding. Therefore, there is a need to create explainable ML or artificial intelligence (XAI) [35]. In our work, we tackle the black-box problem through our design of the descriptors as well as the choice of the type of ML model. Together, we show that our method can distill physics-based insights about the spectral influence of the conductivity of polymers, thereby explaining the source of the model's predictive power.

If the models were to be used for optimization purposes, it is insufficient for them to have good in-distribution/in-dataset accuracy, because the task is to find materials that are better than what is available in the training dataset. Therefore, we also attempt to validate our approach for out-of-distribution, extrapolative performance. In summary, our work aims to achieve the following objectives:

(a) Introduce a two-step workflow to expedite the classification and prediction of highly conductive doped polymer materials. This involves replacing laborious conductivity measurements with highly accurate machine-learned classification and regression models by utilizing UV-VIS-NIR absorbance spectra and bespoke local B-spline descriptors.

(b) Validate the approach's efficacy in classifying and predicting unseen highly conductive samples in an extrapolative manner. This validation is crucial to understand the method's ability to identify new materials with conductivities superior to those present in the dataset.

(c) Present an explainable machine-learned model that offers insights into the influence of absorbance spectra on the conductivities of doped polymer materials.

## 2. Related Work
### 2.1. Different approaches to apply B-splines descriptors

From the view of the method of featurization, B-splines have traditionally been used in curve-fitting [36], numerical differentiation of experimental data [37], computer-aided design [38] and computer graphics [39] due to its useful mathematical properties, e.g. minimal support, completeness with respect to splines of the same order defined on the same knots. In the field of medicine, B-splines based descriptors have been successfully used in classification of the retinal fundus images at an accuracy of 80% [40].

Recently, Fey et. al [41] constructed a spline-based convolutional neural network that was able to improve state-of-the-art results in several benchmark tasks, including image graph classification, graph node classification and shape correspondence on meshes, while allowing very fast training and inference computation. These past works show the range of applicability of B-splines across a wide span of domains.

In a recent work that was closer to ours, Tan et. al [42] showed the use of B-spline, along with Random Forest (RF) and extreme-gradient boosting (XGB) algorithms, to predict the ensemble size and size distribution of gold nanosphere ensembles using domain-informed descriptors of the extinction spectra. The work showed improvement in the performance of the models due to domain-informed descriptors over theory-guided and data-driven descriptors. In contrast, our work focused on purely data-driven featurization within the space of all possible B-splines with the goal to enable high-throughput experimentation and characterization. Unlike their approach, the domain knowledge used in our work only informs that the class of mathematical functions for featurization must be local and positive in character (e.g. B-splines) but does not dictate the specific descriptors themselves. The specific descriptors to be employed within the class were determined in a purely data-driven manner.

## 2.2. The use of absorption spectra in ML

Guda et al., 2021 [43] use X-ray absorption near-edge structure (XANES) spectra and the fingerprint of the local atomic and electronic structures around the absorbing atom. However, they also use handcrafted spectral descriptors, such as edge position, intensities, positions, and curvatures of minima and maxima. Torrisi et al., 2020 [44] use spectral descriptors in random forest models to predict the absorbing atom's local coordination number, mean nearest-neighbor distance, and Bader charge. In addition to point data, their method uses local constant, quadratic and cubic polynomial fits in subdivided domains of the spectra to capture higher-order information, e.g., gradients and curvature. Like our work, the locations of important regions of the spectra were identified. However, the use of random forest models reduces the interpretability of the model. In our work, we show that highly complex random forests and gradient boosting methods are not always necessary. Swapping these for other simpler models, e.g., LASSO (least absolute shrinkage and selection operator), and our B-splines descriptors, would retain the accuracy and further improve the interpretability. Our method not only suggests important spectral regions but also directly indicates the amount each region contributes to the conductivity prediction, and so is easy for interpreting and deriving scientific insights.

Our work also proposes the incorporation of errors in the measurements into the prediction task, which is not the focus of the other works discussed. Intuitively, any learning algorithm should be exposed to the errors of the underlying dataset for it to be able to attenuate or accentuate the amount of weightage given to datapoints of different fidelities. An incorporation of errors would typically improve the performance of the model.

In summary, our work presents a novel approach that incorporates five key elements:(1) a comprehensive experimental dataset containing UV-VIS-NIR spectra of technologically important polymers, (2) a method to engineer B-spline descriptors for spectra in a data-driven fashion, (3) an explainable/interpretable method that leverages mathematical properties of B-spline and LASSO model, (4) a model that takes into account errors in the datapoints, and (5) extrapolative validation of the method.

## 2.2 Explainability

A method is explainable when it is easily understandable by end-users and can provide actionable information that can inform decision making. The notions of explainability that have been explored in literature have been categorized via a hierarchical scheme in a recent review paper [45]. According to the authors, a method's explainability can be judged by human-centred evaluations or objective metric. In this section, we state features of our method that fulfill different identified notions of explainability, and list them in **Table 1**.

Our method has algorithmic transparency and explicitness as it is a linear method with weights that are regularized using $\ell^1$ penalty to select for features. Therefore, the coefficients of the model is sparse and strongly clustered in a spectral region of importance (see **Figure 10**), aiding the interpretation of the relevant absorbance energies that strongly contribute to the conductivity predictions. The causal relationship between higher polaronic densities and higher peaks in the absorbance spectra, which is also a causal factor that determines polymeric conductivity, is captured by the B-spline descriptors that can capture local spectral features to arbitrary precision. Our two-step method is engineered to improve the efficiency over manual process, which supports faster decision-making (see efficiency improvements in **Table 5**). The identified spectral region of importance is corroborated with known HOMO/LUMO gap energies (see **Figure 2(a)**), thus demonstrating its faithfulness to the known opto-physics of the polymers and its justifiability from expert knowledge.

**Table 1** Notions of explainability, their descriptions and feature(s) of our method that fulfills them.

| Notion | Description | Feature of our Method |
|---|---|---|
| Algorithmic Transparency | The degree of confidence of a learning algorithm to behave 'sensibly' in general. | $\ell$1-regularized linear regression |
| Causality | The capacity of a method for explainability to clarify the relationship between input and output | B-spline descriptors that can capture local spectral features to arbitrary precision |

| Efficiency | The capacity of a method for explainability to support faster user decision-making | Efficiency of two-step method (classification and regression models) over manual process |
|---|---|---|
| Explicitness | The capacity of a method to provide immediate and understandable explanations | Sparse and localized coefficients of LASSO Model |
| Faithfulness | The capacity of a method for explainability to select truly relevant features | Large coefficients of B-spline descriptors at spectral regions on the order of HOMO/LUMO gap |
| Justifiability | The capacity of an expert to assess if a model is in line with the domain knowledge | |

[1]Different notions of explainability that are addressed by our method. The notions and their descriptions are extracted from Ref. [45].

## 3. Materials and Methods
### 3.1. Data Acquisition

Electronics grade conjugated polymers P3HT, PBTTT-C12 and PBTTT-C14 were obtained from Reike Metals and Ossilla respectively. The dopant molecules (BCF, F4TCNQ and FeCl3) were purchased from Sigma Aldrich. Both conjugated polymers and dopants were stored in a glovebox where the weighing of the powders and preparation of stock solution were carried out. We used different process solvents (Chlorobenzene, Dichlorobenzene, Chloroform, Tetrahydrofuran, Toluene and Xylene) were used to prepare thin films with varying microstructures.

For solution doping, the stock solution of polymer and dopants were prepared separately and mixed the solution at different volume ratios. The polymer/dopant solution mixture is then spin coated on a quartz substrate of size 2.5 cm x 2.5 cm. The substrates were cleaned using acetone and IPA solution and subsequently underwent ultra-sonification for 10 minutes. The substrates were dried and treated with ozone at a temperature of 100 °C for 10 minutes. The films were prepared by spin coating (spin speed: 1000 rpm and time 90s) and annealed the films at different temperatures. The schematic of the solution doping process is shown in **3.2. (b)**. Solution doping is carried out with BCF as it is soluble in most organic solvents.

For sequential doping, the polymer stock solution in different solvents was prepared and the pristine film was spin coated and annealed at different temperatures before carrying out the doping process. The dopants (F4TCNQ, FeCl3) were dissolved in nitromethane by varying the concentration from 1 mM to 30 mM. The pristine polymer films are then dipped into the dopant solution for ~20 s to achieve complete doping (shown in **3.2. (b)**. The process variables and measurements used in this study are shown in **3.2. (c) & (d)**.

The fabricated films were characterized on the same day in order to avoid the effects of degradation and de-doping. The measurement sequence is the hyperspectral image of the films to get the film uniformity, a four-point probe to measure the sheet resistance, UV-VIS-NIR for spectral descriptors due to polarons and excitons, followed by surface profilometry for thickness measurement.

## 3.2. Data Visualization

A total of 231 samples were synthesized and measured. The maximum conductivity measured is 506 S/cm, minimum is $1.77 \times 10^{-7}$ S/cm, median is 3.87 S/cm. **Figure 3** shows the boxplot for the distribution of the conductivities as a function of dopant type, polymer type, solvent type and annealing temperature (100, 120, 125 and 150º C). A fixed annealing time of 10 minutes was used for all samples. We only performed annealing at temperatures greater than 100 º C for P3HT as the samples for the other polymers were found to be thermally degraded at these temperatures. For each combination of polymer, dopant and annealing time, we varied the amount of dopant and polymer used and recorded the associated dopant and polymer weight ratio (%). See **Table 2** for a complete list of all non-spectral descriptors that were used.

| Type of Descriptors | Name |
|---|---|
| Process Variables | Annealing Temperature (ºC) |
| | Doping Process |
| Material | Polymer Name |
| | Dopant Name |
| | Solvent |
| | Polymer Weight Ratio (%) |
| | Dopant Weight Ratio (%) |
| | Polymer Solution Concentration (mg/ml) |
| | Dopant Solution Concentration (mg/ml) |

| Measurements | Sheet Resistance (Ω/square) |
|---|---|
| | Uniformity (%) |

**Table 2** The non-spectral descriptors which were used in the classification and regression task. There are three types: (1) Process Variables, (2) Material and (3) Measurements.

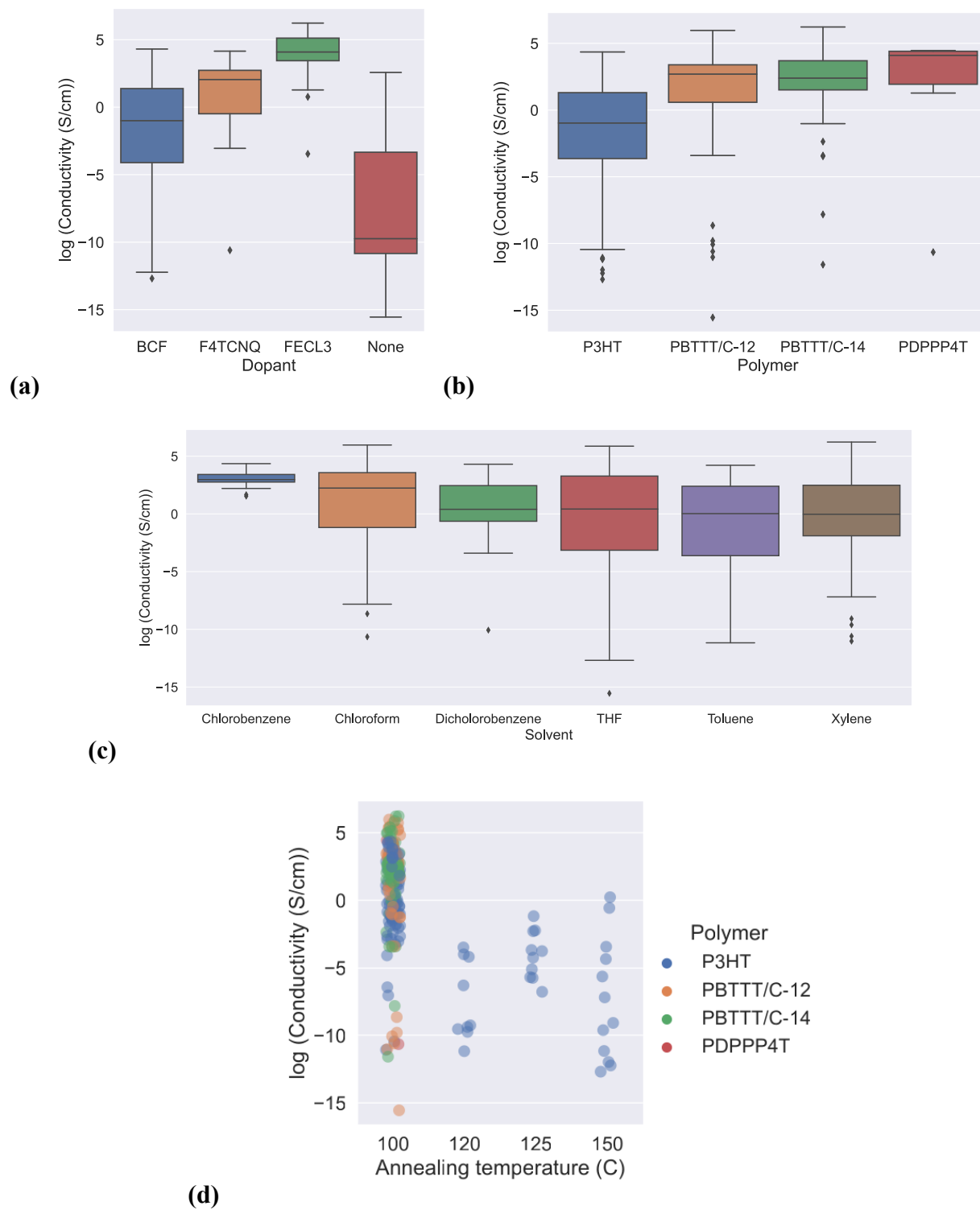

**Figure 3** Boxplots for the distribution of the conductivity of the datapoints as a function of **(a)** dopant type, **(b)** polymer type, **(c)** solvent type and **(d)** annealing temperature.

### 3.3. Data Pre-processing

Besides data related to the process, material and measurements of sheet resistance and uniformity, the UV-VIS-NIR spectrum of each sample was pre-processed. Resampling of the spectral intensity values was performed to regularize the intervals between the datapoints. The mean value of a sub-segment near the left end of the spectra was used to replace the values at the end, which enabled the removal of the spurious sharp left peak. The spectra were smoothed by removing points where the second-order derivative is too large (i.e. greater than 0.0001). Linear interpolation was used to fill in the gaps left by the smoothing procedure. The procedure substantially removes the sharp left peak, which is an unwanted absorption from the OH group of the quartz substrate. Fourier Transform was used to further denoise the spectra by removing frequency components with power spectrum density lower than 0.0001.

The local maxima were found in smoothed and denoised spectra using a peak finding algorithm (find_peak() in SciPy [46]). The inflexion points were found using the same algorithm on the differentiated spectra.

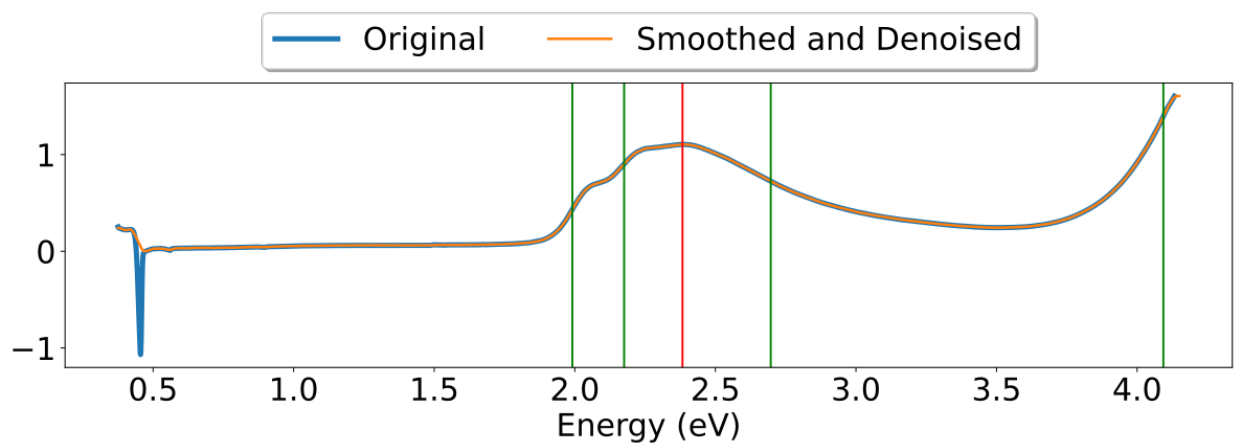

**Figure 4.** The UV-VIS-NIR spectrum of sample *35,* before and after the data pre-processing procedure. The red line denotes the position of the maximum, while the green lines denote the position of inflexions.

### 3.4. Spectral descriptors and ML approach

The goal of supervised ML is to train a model to map from a set of descriptors to a target variable. Spectral, process, material and measurement descriptors were used as inputs. Statistical descriptors of the spectra were used to compare with the proposed B-spline descriptors. In the context of polymer conductivity optimization, the desired outcome is a polymer with higher conductivity. For the classification model, the target is a binary variable which labels a sample to be either highly conductive (1) or not (0). For the regression model, the target is the conductivity of the sample. **Figure 5** shows the schematic of the proposed ML workflow.

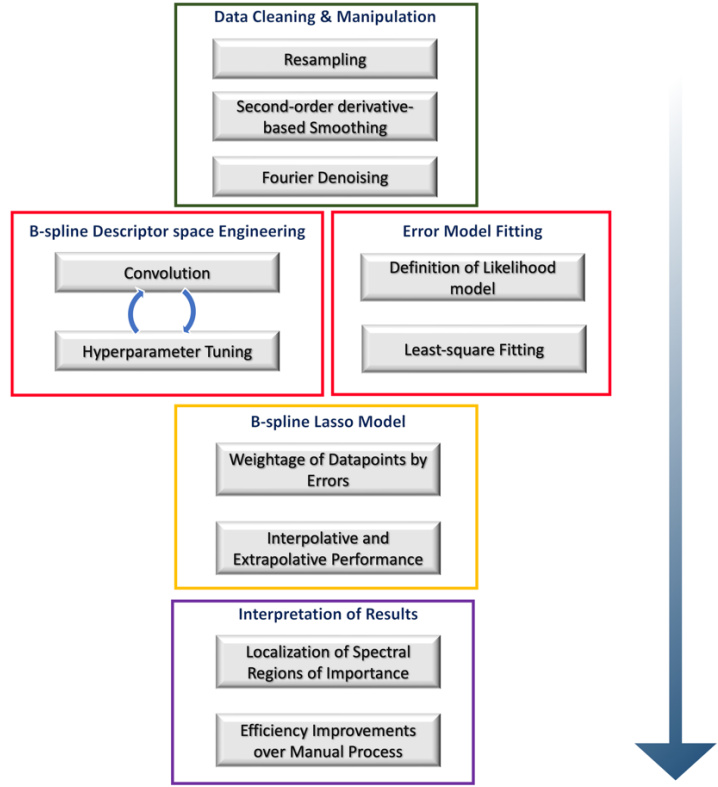

**Figure 5** Schematic of the proposed machine learning workflow.

For the classification task, we minimize the cross-entropy loss,

$$L_{CE} = -\frac{1}{N}\sum_i^N \sum_j^C t_{i,j}\log(f(s_{i,j})) \qquad (1)$$

where

$$f(s_{i,j}) = \frac{e^{s_{i,j}}}{\sum_k^C e^{s_{i,k}}} \qquad (2)$$

And $s_{i,j}$ represents predictions by the model of either the two classes (i.e. highly conductive samples, and those which are not, C=2) and $t_{i,j}$ is an indicator variable for the correct classification of observation i.

For the regression task, we employ the MSE loss,

$$L_{MSE} = \frac{1}{N}\sum_{i=1}^N (y_i - \hat{y}_i)^2 \qquad (3)$$

Where $y_i$ represents the predictions by the model of the target conductivity and $\hat{y}_i$ is the ground-truth of the conductivity.

## 3.5 Spectral descriptors

### 3.5.1. Statistical Descriptors

There are many ways to generate point spectral data. To identify the most performant definition as our baseline, we conducted preliminary tests on 5 different types of point spectral data by training a decision tree classification model to identify samples with high conductivity. We define high conductivity as samples with log(Conductivity) > 0.1 as this results in around 53 out of 231 datapoints (~23%) being in the high conductivity class, giving enough samples for the model to learn the right mapping. The 5 types of point spectral data are: (1) regularly spaced points, (2) curated points by a domain expert, (3) regularly spaced points on first-order differentiated spectra, (4) curated points on first-order differentiated spectra by a domain expert and (5) statistics of maxima and inflexions. For (2) and (4), regions of the spectra that could inform the conductivity task were chosen by one of the authors who is an organic chemistry expert. The density of points for (1) and (3) were tuned, along with the maximum depth of the decision tree in a 3-fold 50:50 stratified cross-validation setting.

The best result of the tests suggests that the statistics of the maxima and inflexions were the most performant point data, with a test accuracy of 90%, while the others were at 70% or lower. Therefore, we use them to compare with our proposed B-splines descriptors as discussed below.

### 3.5.2. B-spline Descriptors

We define our bespoke B-spline descriptors for the polymer absorbance spectra in this section. It is well-known that the optical physics of polymer systems is related to the intensity and position of spectral descriptors, such as peaks, troughs and shoulders, of the polymer systems. Therefore, the electrical conductivity of the systems would also be correlated with the spectra, as both are derived from the electronic response to electromagnetic perturbation. In particular, the semiconducting nature of such polymers would suggest that there are distinguishable phenomena that are triggered at specific electromagnetic (EM) radiation energy on the scale of the polymer HOMO/LUMO gap. Together, they suggest that descriptors that are local would be more performant than global ones.

In view of this, we propose B-splines to generate a good set of descriptors for the conductivity prediction task. B-splines are a kind of function made up of polynomial pieces, which are useful for approximating other functions. They have been shown to require the least amount of support for a given degree, smoothness, and domain partition. A spline function of degree $n$ is a polynomial function with degree *n-1* that is piecewise defined in terms of a variable *x*, and the points at which the pieces connect are known as knots. Spline functions have a noteworthy characteristic that they and their derivatives are continuous, depending on the number of times each knot appears. B-splines of order *n* serve as a foundation for spline functions of the same order that are defined over the same knots. This means that any possible spline function can be created by combining B-splines linearly, and each spline function has a unique combination.

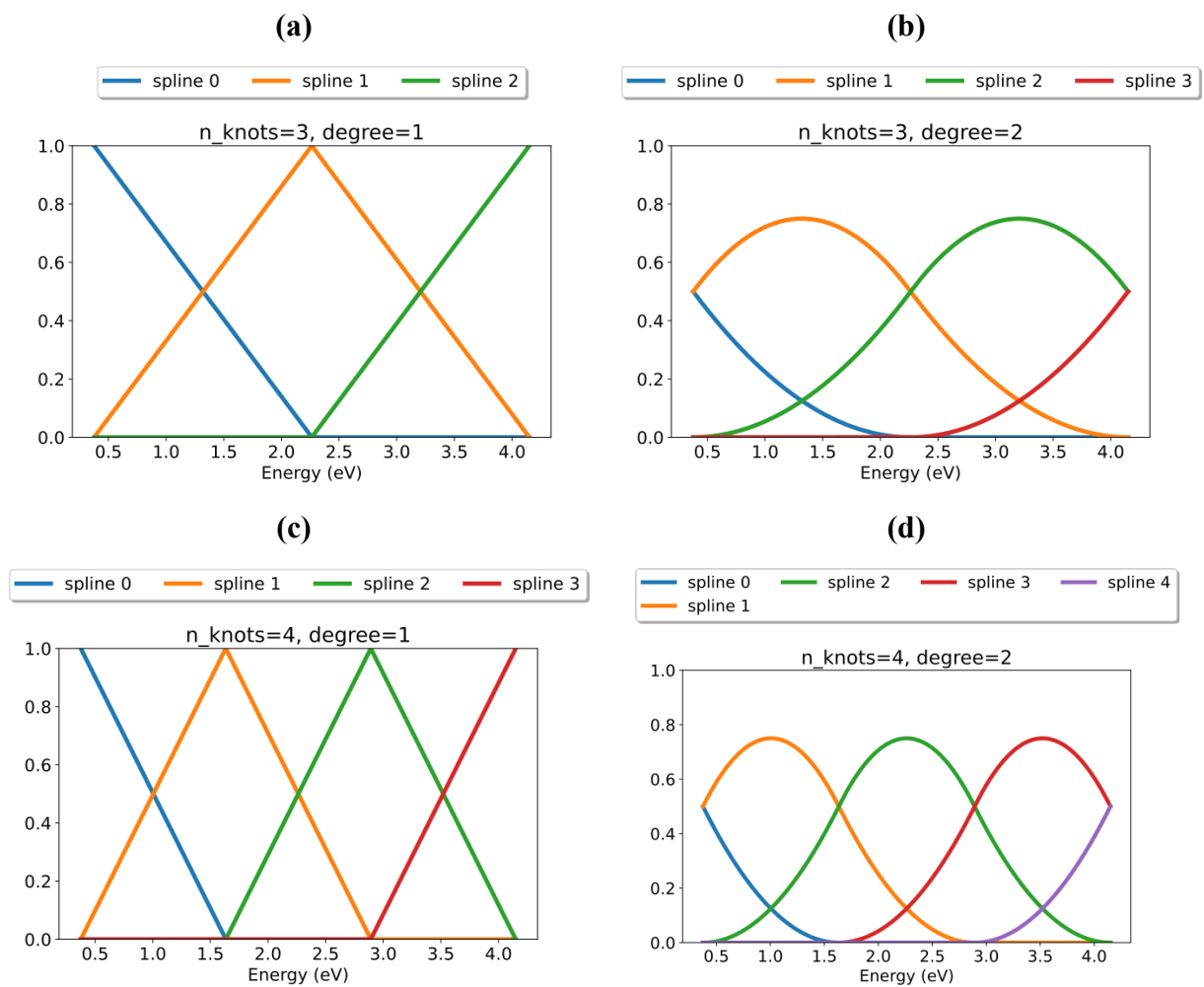

**Figure 6** Illustrations of B-splines of different numbers of uniformly spaced knots (n_knots) and degree as a function of the independent quantity of the absorbance spectra (i.e., energy).

Due to the mentioned compactness, uniqueness, and positive nature of the B-splines, we deemed them suitable to be used as the basis functions to generate spectral descriptors. The compactness and uniqueness will ensure that the spectral information is robustly represented by the descriptors. The positive nature of both the B-splines and the absorbance spectra will guarantee the easy interpretation of the spectral descriptors, which we define as the projection of the spectra onto the B-splines via a convolutional operation, as they are guaranteed to be only positive quantities.

Mathematically, B-splines can be defined by means of the Cox–de Boor[47] recursion formula. Given a knot sequence …, $t_0$, $t_1$, $t_2$, …, the B-splines of order 1 are defined by,

$$B_{i,1}(x) := \begin{cases} 1 & if\ t_i<x<t_{i+1} \\ 0 & otherwise \end{cases} \quad (4)$$

The higher-order B-splines are defined by recursion,

$$B_{i,k+1}(x) := \omega_{i,k}(x)B_{i,k}(x) + (1 - \omega_{i+1,k}(x))B_{i+1,k}(x) \quad (5)$$

where

$$\omega_{i,k}(x) := \begin{cases} \frac{x-t_i}{t_{i+k}-t_i}, & t_{i+k} \neq t_i \\ 0 & otherwise \end{cases} \quad (6)$$

The B-spline descriptors, $d_{i,k}$ of a spectrum $S(x)$ are defined through a convolutional operation,

$$d_{i,k} = \int B_{i,k}(x) * S(x)\, dx \quad (7)$$

In our case, we have chosen to use knot sequence with equal spacing. The other hyperparameters of the B-spline descriptors were determined using a K-fold cross-validated hyperparameter tuning method. To elaborate, the B-spline descriptors, $d_{i,k}$, are projections of the absorbance spectrum $S(x)$ onto the B-splines, $B_{i,k}$, and are local in nature.

The descriptive power of the choice of B-spline descriptors was tested by fitting the spectra with a ridge regressor. **Figure 7** shows the behavior of the ridge regression models produced by fitting the B-spline descriptors with the varying number of knots and degree of the polynomial. There is an observed change of the fit from having sharp edges to smooth edges as the degree goes from 1 (linear) to 2 (quadratic) (i.e., left column to right column). Also, the accuracy of the fit improves with the number of knots going from 10 to 20 (i.e., top row vs bottom row). The good accuracy of the fit for **Figure 7(d)**, where most of the oscillating features of the spectrum are increasingly accurately described, suggests that the B-splines have the ability to model very complicated local features of the spectra, which is crucial for our prediction task.

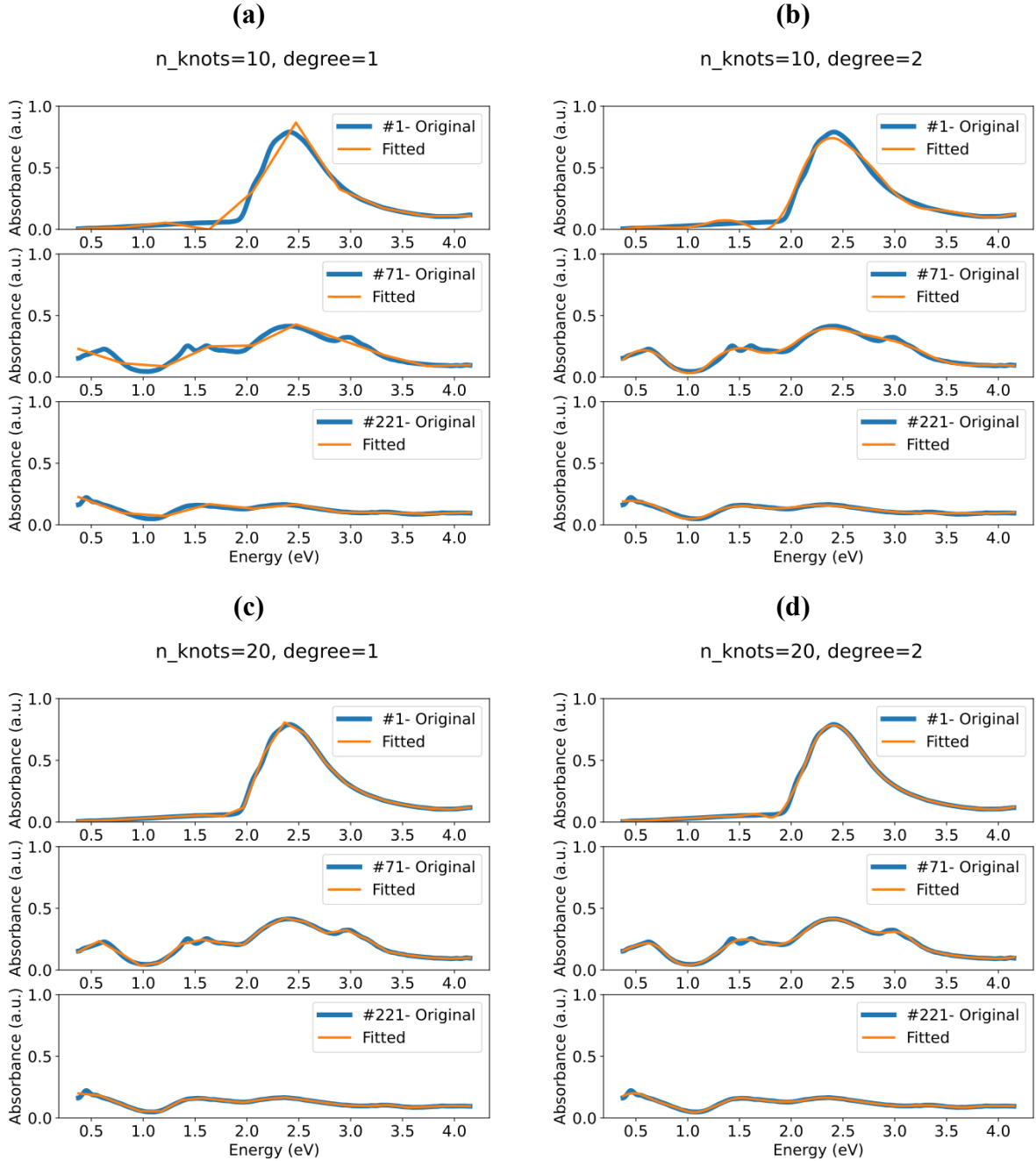

**Figure 7** The original and fitted spectra of 3 samples (#1, #71 and #221) to illustrate the descriptive power of the B-spline descriptors for describing the spectra. **(a)** number of knots = 10 and the degree of the polynomial = 1, **(b)** number of knots = 10 and the degree of the polynomial = 2, **(c)** number of knots = 20 and the degree of the polynomial = 1 and **(d)** number of knots = 20 and the degree of the polynomial = 2.

Theoretically, the statistical descriptors only contain point-wise information while the B-spline descriptors can discern higher-order information of the spectra, e.g., the local slope and curvature. Empirically, when put through the same hyperparameter tuning procedure, the most performant statistical descriptors perform not as good as B-spline descriptors in terms of $R^2$ and MAE metrics (see Results and Discussions).

### 3.5.3 Experimental and Measurement Descriptors

The process variables, materials and measurements of sheet resistance and uniformity were included during the training of the classification and regression model.

## 3.6 Classification and Regression Models

Regression models for polymer conductivity were trained using spectral, experimental and measurement descriptors. Specifically, we target the $\log_{10}$(conductivity) as the raw conductivity values span many orders of magnitude. For spectral descriptors, we conducted separate runs for statistical and B-spline descriptors to investigate their separate performances. We used the coefficient of $R^2$ to measure the performance of the models. The best possible score is 1.0, and the coefficient can be negative.

For all training runs, we conducted stratified sampling using the criterion of $\log_{10}$(Conductivity) > 0.1 to divide the dataset into high and low conductive samples. This is to ensure that the model have enough statistics of both low and high conductive samples to distinguish between them, as good model performance for predicting samples of high conduction is needed for the optimization task. It is a critical step for the small dataset because a random sample may produce a disproportionately small number of samples of highly conductive polymers in either the train or test fit, thereby reducing the quality of the chosen performance metric in measuring the true performance of the model. We chose a train-test split ratio of 90:10.

The hyperparameter search was conducted with 15-fold stratified cross-validation and deemed sufficient as any increase in the number of folds would only result in small changes in validation $R^2$. Hyperparameters that were tuned (stated in their argument name in Scikit-learn package [48]): (1) Lasso – alpha, (2) Random Forest - n_estimators, max_depth, (3) Gradient Boosting – n_estimators, max_depth, (4) B-spline descriptors – n_knots. The third-degree polynomial was used in every case to generate the B-spline descriptors as we deemed it flexible enough to model the small fluctuations in the spectra.

The classification model used in our work to distinguish highly conductive samples from the others is directly obtained from the regression model by the choice of an appropriate conductivity cutoff.

### 3.7 Error model

The errors of datapoints should inform the relative weightage given to them during the training process. One should expect an improvement in performance due to the relatively less weight given to erroneous points. One straightforward way to incorporate errors in regression tasks is to interpret the training procedure as performing a maximum likelihood procedure. In particular, the likelihood for a linear model, y = ax + b, for a set of datapoints $\{x_i, y_i\}$ with different uncertainties/variances $\{\sigma_i\}$ is given by,

$$L \propto \prod e^{-\frac{1}{2}\left(\frac{y_i - (ax_i + b)}{\sigma_i}\right)^2} \qquad (5)$$

Therefore, the weightage factor to be associated with each datapoint is $1/\sigma_i^2$. This equation applies directly to the LASSO model used in this work.

The conductivities of the polymer samples were measured up to three times. Since we want to predict the $\log_{10}$(conductivity) of the samples, we need its associated errors. There are 49 samples (of a total of 231) with 2 or 3 measurements of conductivity. The rest (182) have only a single conductivity measurement. To estimate the errors of the $\log_{10}$(conductivity) of the latter group, we fitted a linear model to the logarithm of the standard deviation of the $\log_{10}$(conductivity) of the former group with respect to their mean of $\log_{10}$(conductivity). The model for the errors is then used to predict the latter group's standard deviation. **Figure 8** shows the standard deviation and mean of $\log_{10}$(conductivity) for all points. The figure shows many of the red datapoints lie outside the range of the blue points, indicating the extrapolative nature of these predictions, hence we did not attempt to apply more complicated regression models, which would tend to have larger errors for extrapolative predictions. We continue our use of the error model and its predictions in our further analyses while being mindful of the discussed deficiencies.

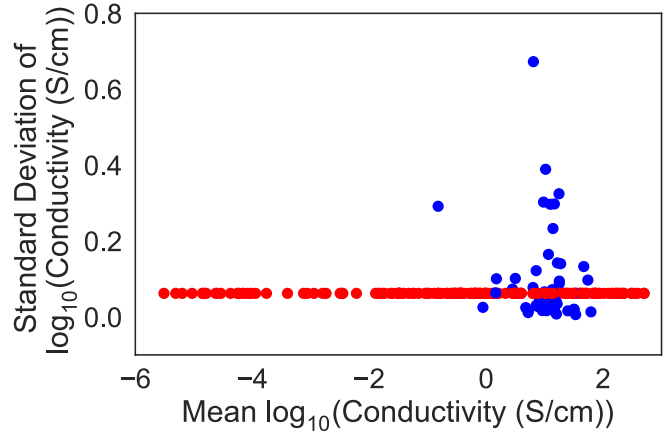

**Figure 8** Standard deviation of $\log_{10}$(Conductivity) against the mean of $\log_{10}$(Conductivity) for both sets of samples with 2 or 3 conductivity measurements (blue points) and those with only 1 (red points). The standard deviations of the red points are predicted from a linear regression model fitted onto the blue.

## 4 Results and Discussions

**Table 3** shows the performance comparison between the statistical descriptors and our proposed B-spline descriptors, both fitted along with the measurement and experimental descriptors. Both training and testing $R^2$ of the B-spline descriptors are better than the statistical descriptors, for both Random Forest and Gradient Boosting methods, indicating the advantage of using our proposed descriptors.

| Model | Statistical Descriptors | | B-spline Descriptors | |
|---|---|---|---|---|
| | **Train $R^2$** | **Test $R^2$** | **Train $R^2$** | **Test $R^2$** |
| **Random Forest** | 0.923 | 0.804±0.035 | **0.992** | **0.983±0.025** |
| **Gradient Boosting** | 0.978 | 0.775±0.041 | **0.993** | **0.983±0.024** |

**Table 3** Comparison of training and testing $R^2$ for models using either statistical descriptors or B-spline descriptors. The descriptors were tested on random forest and gradient boosting models. The best of the train and test $R^2$ are in bold.

To test the interpretability of our B-spline descriptors, we fitted them without the experimental and measurement descriptors to the conductivity using a LASSO model. **Figure 9** shows the true positive rate and false positive rate of using the regression model against various $\log_{10}$(Conductivity) cutoffs. Interestingly, we identified a region (shaded in grey) that corresponds to a classification accuracy of 100%. The plot of the coefficient of the B-spline descriptors of the

LASSO model at different points along the energy axis shows large magnitudes around the 1.5-2.0 eV region (**Figure 10**). This is expected as the LASSO model tends to perform descriptor selection and suppresses uninformative descriptors. It is known that the electronic physics of such polymer systems depends on dopant levels, which lie within the HOMO/LUMO gap. The spectral region of importance corroborates well with the known typical scale of HOMO/LUMO energy in such polymer systems of about 1.45-2.0 eV (see **Figure 2 (a)**). We note that such causal information was not handcrafted into the model but learned in a data-driven manner by the model, removing the need for expert knowledge. Furthermore, the guaranteed positive nature of the B-spline descriptors, as discussed previously, means that the value of the coefficients of the LASSO model directly predict the local spectral contributions to the conductivity prediction. The shape of the coefficient plot in **Figure 10** indicates that a spectrum with a peak between two troughs around 1.5-2.0 eV would result in high conductivity. This level of interpretability is arguably not possible when other types of descriptors are used with other more complicated models.

The advantage of converting a regression model into a classification model is that one would not need to decide the class identity of samples a priori. This means that one can, depending on the regression model's characteristics, decide upon the appropriate tradeoff between sensitivity and specificity.

The accuracy (Acc), sensitivity, specificity [49] of a classifier are defined as:

$$Acc = \frac{TP+TN}{TP+TN+FP+FN} \tag{6}$$

$$Sensitivity = \frac{TP}{TP+FN} \tag{7}$$

$$Specificity = \frac{TN}{TN+FP} \tag{8}$$

where TP, TN, FP, and FN stand for true positive, true negative, false positive and false negative, respectively.

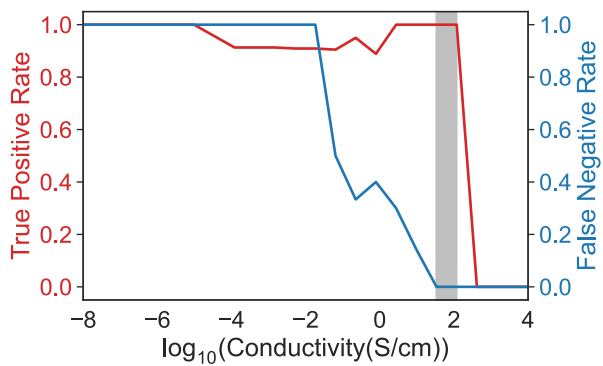 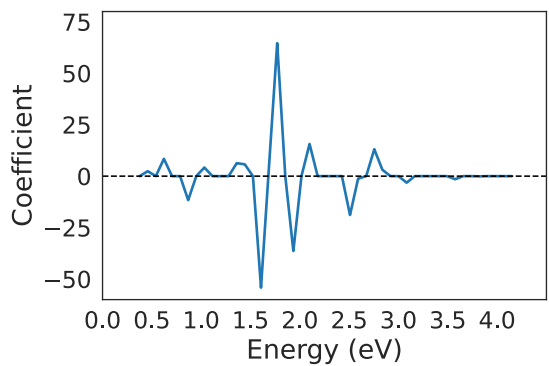

**Figure 9** True positive rate and false negative rate of a LASSO classification model for log (Conductivity) greater than different values. The shaded region for conductivities from approximately $10^{1.4}$ to $10^2$ S/cm has an accuracy of 100% for the classification task of identifying high conductivity samples.

**Figure 10** The coefficient of the B-spline basis functions used in the classification model across the whole absorbance spectrum. Notice the sparsity of the model with respect to the basis functions.

It could be argued that the accuracy of our model has not been maximized because LASSO models are very simple. To investigate if a more complicated models may yield better performance, we conduct similar runs with hyperparameter tuning for random forest and gradient boosting models as well. For this training run, we included the experimental and measurement descriptors to investigate the performance when all information is included. Furthermore, we trained a LASSO model with the appropriate weights, either empirically determined or given by our error model, to account for datapoint errors. Beyond measuring the performance of our model with the usual train and test $R^2$ (**Figure 11(a)**), we devised a test for extrapolative abilities. In optimization settings, where one is aiming to achieve materials of better properties than those in the dataset, it is crucial to test a model's extrapolative abilities. To do so, we excluded the top-2 highest conductivity samples from the dataset and tested specifically for the extrapolative performance using them. The top-2 highest conductivities are at least 29% higher than the rest of the dataset (**Figure 11(b)**).

**Table 4** summarizes the various metrics for the four models. We observe that though LASSO models have lower train $R^2$, they have larger test $R^2$ than the other two more complicated models, even when the hyperparameters of all models were tuned. The mean absolute error of the top-2 extrapolative datapoints of LASSO models is less than half of the random forest and gradient boosting methods. In total, LASSO models have better general interpolative and extrapolative performance than more complicated methods.

**Table 4** The train, test $R^2$ metric and the mean absolute top-2 extrapolation errors for Lasso, Random Forest and Gradient Boosting methods. For Lasso, the models with and without errors are included. The best of the train and test $R^2$, as well as the mean absolute Top-2 extrapolation error, are in bold.

| Dataset | Algorithms | Train $R^2$ | Test $R^2$ | Train MAE | Test MAE | Mean Absolute Top-2 Extrapolation Error (S/cm) |
|---|---|---|---|---|---|---|
| P3HT | LASSO (Using Error) | 0.990 ±0.001 | **0.986 ±0.012** | 0.125 ±0.003 | **0.163 ±0.041** | 25 |
|  | LASSO (No Error) | 0.965 ±0.007 | 0.932 ±0.088 | 0.193 ±0.014 | 0.221 ±0.100 | **14** |
|  | RF | 0.995 ±0.001 | 0.968 ±0.025 | 0.093 ±0.003 | 0.214 ±0.053 | 37 |
|  | GB | 0.999 ±0.000 | 0.982 ±0.012 | 0.012 ±0.001 | 0.184 ±0.033 | 29 |
|  | MLP | 0.989 ±0.003 | 0.915 ±0.100 | 0.126 ±0.025 | 0.315 ±0.184 | 29 |
| PBTTT/C-12 | LASSO (Using Error) | 0.998 ±0.000 | 0.965 ±0.030 | 0.166 ±0.006 | 0.148 ±0.070 | 174 |
|  | LASSO (No Error) | 0.998 ±0.000 | **0.981 ±0.048** | 0.124 ±0.005 | **0.126 ±0.051** | **163** |
|  | RF | 0.994 ±0.035 | 0.968 ±0.001 | 0.194 ±0.027 | 0.211 ±0.029 | 201 |
|  | GB | 0.995 ±0.004 | 0.959 ±0.019 | 0.110 ±0.043 | 0.182 ±0.025 | 212 |
|  | MLP | 0.990 ±0.097 | 0.914 ±0.068 | 0.126 ±0.178 | 0.318 ±0.102 | 225 |
| PBTTT/C-14 | LASSO (Using Error) | 0.982 ±0.017 | **0.984 ±0.007** | 0.134 ±0.012 | **0.156 ±0.012** | **112** |
|  | LASSO (No Error) | 0.974 ±0.026 | 0.906 ±0.032 | 0.196 ±0.075 | 0.258 ±0.071 | 134 |
| | RF | 0.988 ±0.028 | 0.973 ±0.003 | 0.135 ±0.053 | 0.213 ±0.010 | 182 |
| | GB | 0.990 ±0.044 | 0.957 ±0.049 | 0.018 ±0.014 | 0.234 ±0.029 | 191 |
| | MLP | 0.988 ±0.157 | 0.912 ±0.089 | 0.093 ±0.079 | 0.289 ±0.083 | 173 |
| PDPPP4T | LASSO (Using Error) | 0.999 ±0.002 | **0.981** ±0.050 | 0.127 ±0.006 | **0.168** ±0.012 | **156** |
| | LASSO (No Error) | 0.962 ±0.029 | 0.968 ±0.009 | 0.194 ±0.031 | 0.218 ±0.006 | 166 |
| | RF | 0.996 ±0.058 | 0.981 ±0.003 | 0.096 ±0.013 | 0.181 ±0.013 | 210 |
| | GB | 0.990 ±0.040 | 0.968 ±0.034 | 0.013 ±0.018 | 0.181 ±0.002 | 222 |
| | MLP | 0.990 ±0.056 | 0.915 ±0.153 | 0.126 ±0.071 | 0.316 ±0.158 | 210 |
| Multi-polymers | LASSO (Using Error) | 0.977 ±0.003 | **0.972** ±0.027 | 0.175 ±0.006 | 0.187 ±0.046 | **139** |
| | LASSO (No Error) | 0.978 ±0.003 | 0.972 ±0.028 | 0.170 ±0.007 | **0.189** ±0.048 | 141 |
| | RF | 0.994 ±0.001 | 0.963 ±0.023 | 0.107 ±0.004 | 0.227 ±0.033 | 312 |
| | GB | 0.999 ±0.000 | 0.970 ±0.021 | 0.055 ±0.002 | 0.204 ±0.057 | 332 |
| | MLP | 0.981 ±0.004 | 0.919 ±0.042 | 0.169 ±0.020 | 0.338 ±0.076 | 297 |

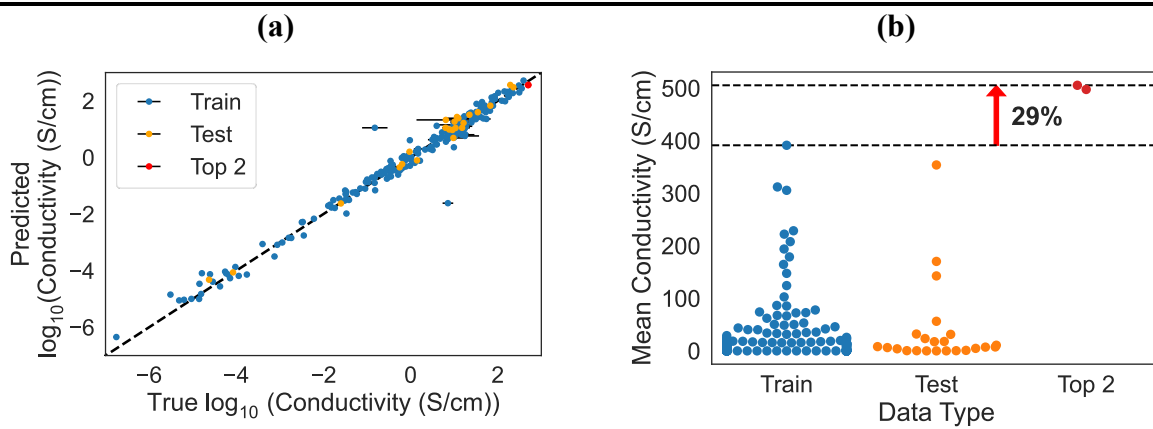

**Figure 11 (a)** Parity plot of the LASSO model with errors for various splits of the data. The errors of the conductivity are shown as horizontal bars, **(b)** the distribution of mean conductivity of the various splits of the data.

One of the goals of our work is to investigate the potential efficiency improvement in laborious tasks using classification and regression models. The classifier could be used to downsample the number of samples to characterize manually at close to perfect accuracy. The number of such samples would be affected by the choice of conductivity threshold. If a high conductivity threshold is chosen, the number of samples suggested by the classifier would be low, but one runs the risk of missing some actual high conductivity samples, and vice versa. **Table 5** shows the various efficiency improvement as a percentage of the total time with a conservative ($10^{1.4}$ S/cm) or less conservative ($10^2$ S/cm) choice of threshold, where the classification accuracy is found to be 100%. After the classification, one can choose either (1) to conduct actual measurements of thickness and sheet resistance, or (2) proceed to conduct a UV-VIS-NIR spectral imaging and use the data in the regression model, to predict the conductivity. In our analyses, we assumed, from our experience, 3, 2 and 3 minutes to conduct thickness, sheet resistance and UV-VIS-NIR measurements, respectively. If we use other high throughput spectroscopic techniques, e.g. hyperspectral imaging, we can reduce the time taken significantly. We see that conservatively one could expect a 25% efficiency improvement if we opt for a lower threshold and direct measurement. When using a higher threshold and the regression model, our method achieves the best efficiency improvement of 59%, which is 89% of the theoretical maximum using our experimental setup. Refer to the supplementary section for more information about the calculations. We emphasize that using hyperspectral imaging techniques, the time for imaging could be drastically reduced resulting in the further efficiency gain. We note that a higher degree of down-sampling could be achieved with the classifier using a higher threshold, resulting in better measurement efficiency improvement, but at the cost of lower classification accuracy.

|  |  | Efficiency Improvement (%) | |
|---|---|---|---|
|  |  | Second-step Method | |
|  |  | Direct Measurement | B-spline LASSO Regressor |
| Threshold of First-step Classifier (S/cm) | $10^{1.4}$ | 25 | 47 |
|  | $10^2$ | 48 | 59 |
| Theoretical Maximum |  | 67 | |

**Table 5** Efficiency gains using different first-step and second-step methods by comparing the experimentation times.

## 5 Limitations

The major limitation of this approach is that the correlation between the low energy spectral features to electrical conductivity is not captured due to low signal strength. In addition, the presence of artifacts due to substrates or other contaminants might produce wrong prediction. Therefore, it is important to pre-process the spectral data before machine learning. The absorption spectra of extended infrared wavelength range might provide added advantage of capturing all the relevant spectral features corresponding to electrical conductivity that will further increase the accuracy of the proposed method.

## 6 Conclusions

We proposed a two-step workflow which involves a data-driven ML approach using a pair of classification and regression models, taking into account measurement errors/uncertainties, to reduce the need for manual intervention in conductivity measurements of doped conjugated polymer systems. The proposed data-driven local B-spline descriptors outperform the best statistical descriptors for the conductivity regression task. The B-spline LASSO regression model has a test $R^2$ of 0.984, outperforming both the Random Forest and Gradient Boosting models. The classification model, derived from the regression model, has an accuracy of 100% for conductivity $>\sim$ 25 to 100 S/cm. Validation of our method was conducted by predicting the samples with the two highest conductivities in an extrapolative manner - the Mean Absolute Top-

2 Extrapolation Error for our model is less than half of the other two. The proposed workflow results in an improvement in measurement efficiency by 59%, which is 89% of the maximum achievable. We showed how our approach resulted in a set of descriptors and a model that was able to learn to use causally relevant spectral information in a purely data-driven manner. This study offers a way to accelerate the optimization of doped polymer materials and demonstrates the insights that can be gained with deliberate uses of ML in experimental science.

## Acknowledgements

This study was supported by the Accelerated Materials Development for Manufacturing Program at A*STAR via the AME Programmatic Fund by the Agency for Science, Technology and Research, Singapore under Grant No. A1898b0043.

**CRediT authorship contribution statement:**

**Ji Wei Yoon**: Conceptualization, Methodology, Software, Data Curation, Validation, Visualization, Formal analysis, Writing- Original Draft, Writing - Review & Editing. **Adithya Kumar:** Investigation, Data Curation, Writing - Review & Editing. **Pawan Kumar:** Investigation, Writing - Review & Editing. **Kedar Hippalgaonkar**: Resources, Supervision, Project Administration, Funding acquisition, Writing - Review & Editing. **J Senthilnath:** Formal analysis, Funding acquisition, Resources, Project Administration, Writing - Review & Editing, Supervision. **Chellappan Vijila:** Conceptualization, Supervision, Investigation, Visualization, Writing- Original Draft, Writing - Review & Editing.

# References

[1] R. Gross, R. Hanna, A. Gambhir, P. Heptonstall, J. Speirs, How long does innovation and commercialisation in the energy sectors take? Historical case studies of the timescale from invention to widespread commercialisation in energy supply and end use technology, Energy Policy. 123 (2018) 682–699. https://doi.org/10.1016/J.ENPOL.2018.08.061.

[2] P.M. Attia, A. Grover, N. Jin, K.A. Severson, T.M. Markov, Y.H. Liao, M.H. Chen, B. Cheong, N. Perkins, Z. Yang, P.K. Herring, M. Aykol, S.J. Harris, R.D. Braatz, S. Ermon, W.C. Chueh, Closed-loop optimization of fast-charging protocols for batteries with machine learning, Nature 2020 578:7795. 578 (2020) 397–402. https://doi.org/10.1038/s41586-020-1994-5.

[3] S. Langner, F. Häse, J.D. Perea, T. Stubhan, J. Hauch, L.M. Roch, T. Heumueller, A. Aspuru-Guzik, C.J. Brabec, Beyond Ternary OPV: High-Throughput Experimentation and Self-Driving Laboratories Optimize Multicomponent Systems, Advanced Materials. 32 (2020) 1907801. https://doi.org/10.1002/ADMA.201907801.

[4] V. Chaudhary, R. Chaudhary, R. Banerjee, R. V. Ramanujan, Accelerated and conventional development of magnetic high entropy alloys, Materials Today. 49 (2021) 231–252. https://doi.org/10.1016/J.MATTOD.2021.03.018.

[5] Y. Zuo, M. Qin, C. Chen, W. Ye, X. Li, J. Luo, S.P. Ong, Accelerating materials discovery with Bayesian optimization and graph deep learning, Materials Today. 51 (2021) 126–135. https://doi.org/10.1016/J.MATTOD.2021.08.012.

[6] B. Tang, Y. Lu, J. Zhou, T. Chouhan, H. Wang, P. Golani, M. Xu, Q. Xu, C. Guan, Z. Liu, Machine learning-guided synthesis of advanced inorganic materials, Materials Today. 41 (2020) 72–80. https://doi.org/10.1016/J.MATTOD.2020.06.010.

[7] F. Akhundova, L. Lüer, A. Osvet, J. Hauch, I.M. Peters, K. Forberich, N. Li, C. Brabec, Building process design rules for microstructure control in wide-bandgap mixed halide perovskite solar cells by a high-throughput approach, Appl Phys Lett. 118 (2021) 243903. https://doi.org/10.1063/5.0049010.

[8] B.P. MacLeod, F.G.L. Parlane, C.C. Rupnow, K.E. Dettelbach, M.S. Elliott, T.D. Morrissey, T.H. Haley, O. Proskurin, M.B. Rooney, N. Taherimakhsousi, D.J. Dvorak, H.N. Chiu, C.E.B. Waizenegger, K. Ocean, M. Mokhtari, C.P. Berlinguette, A self-driving laboratory advances the Pareto front for material properties, Nature Communications 2022 13:1. 13 (2022) 1–10. https://doi.org/10.1038/s41467-022-28580-6.

[9] A. Abutaha, V. Chellappan, P. Kumar, K. Hippalgaonkar, Linking Polaron Signatures to Charge Transport in Doped Thiophene Polymers, ACS Appl Energy Mater. (2023). https://doi.org/10.1021/ACSAEM.3C00149/ASSET/IMAGES/MEDIUM/AE3C00149_0006.GIF.

[10] A. Abtahi, S. Johnson, S.M. Park, X. Luo, Z. Liang, J. Mei, K.R. Graham, Designing π-conjugated polymer blends with improved thermoelectric power factors, J Mater Chem A Mater. 7 (2019) 19774–19785. https://doi.org/10.1039/C9TA07464C.

[11] P.Y. Yee, D.T. Scholes, B.J. Schwartz, S.H. Tolbert, Dopant-Induced Ordering of Amorphous Regions in Regiorandom P3HT, Journal of Physical Chemistry Letters. 10 (2019) 4929–4934. https://doi.org/10.1021/ACS.JPCLETT.9B02070/SUPPL_FILE/JZ9B02070_SI_001.PDF.

[12] S.N. Patel, A.M. Glaudell, K.A. Peterson, E.M. Thomas, K.A. O'Hara, E. Lim, M.L. Chabinyc, Morphology controls the thermoelectric power factor of a doped

semiconducting polymer, Sci Adv. 3 (2017). https://doi.org/10.1126/SCIADV.1700434/SUPPL_FILE/1700434_SM.PDF.

[13] J. Hynynen, D. Kiefer, C. Müller, Influence of crystallinity on the thermoelectric power factor of P3HT vapour-doped with F4TCNQ, RSC Adv. 8 (2018) 1593–1599. https://doi.org/10.1039/C7RA11912G.

[14] K. Namsheer, C.S. Rout, Conducting polymers: a comprehensive review on recent advances in synthesis, properties and applications, RSC Adv. 11 (2021) 5659–5697. https://doi.org/10.1039/D0RA07800J.

[15] K.K. Sadasivuni, S.M. Hegazy, Aa.A.M. Abdullah Aly, S. Waseem, K. Karthik, Polymers in electronics, Polymer Science and Innovative Applications: Materials, Techniques, and Future Developments. (2020) 365–392. https://doi.org/10.1016/B978-0-12-816808-0.00011-1.

[16] S.N. Patel, A.M. Glaudell, K.A. Peterson, E.M. Thomas, K.A. O'Hara, E. Lim, M.L. Chabinyc, Morphology controls the thermoelectric power factor of a doped semiconducting polymer, Sci Adv. 3 (2017). https://doi.org/10.1126/SCIADV.1700434/SUPPL_FILE/1700434_SM.PDF.

[17] J. Saska, G. Gonel, Z.I. Bedolla-Valdez, S.D. Aronow, N.E. Shevchenko, A.S. Dudnik, A.J. Moulé, M. Mascal, A Freely Soluble, High Electron Affinity Molecular Dopant for Solution Processing of Organic Semiconductors, Chemistry of Materials. 31 (2019) 1500–1506. https://doi.org/10.1021/ACS.CHEMMATER.8B04150/SUPPL_FILE/CM8B04150_SI_001.PDF.

[18] V. Untilova, T. Biskup, L. Biniek, V. Vijayakumar, M. Brinkmann, Control of Chain Alignment and Crystallization Helps Enhance Charge Conductivities and Thermoelectric Power Factors in Sequentially Doped P3HT:F4TCNQ Films, Macromolecules. 53 (2020) 2441–2453. https://doi.org/10.1021/ACS.MACROMOL.9B02389/SUPPL_FILE/MA9B02389_SI_001.PDF.

[19] V. Vijayakumar, P. Durand, H. Zeng, V. Untilova, L. Herrmann, P. Algayer, N. Leclerc, M. Brinkmann, Influence of dopant size and doping method on the structure and thermoelectric properties of PBTTT films doped with F6TCNNQ and F4TCNQ, J Mater Chem C Mater. 8 (2020) 16470–16482. https://doi.org/10.1039/D0TC02828B.

[20] A. Shokry, M. Karim, M. Khalil, S. Ebrahim, J. El Nady, Supercapacitor based on polymeric binary composite of polythiophene and single-walled carbon nanotubes, Scientific Reports 2022 12:1. 12 (2022) 1–13. https://doi.org/10.1038/s41598-022-15477-z.

[21] J.E. Cochran, M.J.N. Junk, A.M. Glaudell, P.L. Miller, J.S. Cowart, M.F. Toney, C.J. Hawker, B.F. Chmelka, M.L. Chabinyc, Molecular interactions and ordering in electrically doped polymers: Blends of PBTTT and F4TCNQ, Macromolecules. 47 (2014) 6836–6846. https://doi.org/10.1021/MA501547H/SUPPL_FILE/MA501547H_SI_001.PDF.

[22] K. Liu, B. Ouyang, X. Guo, Y. Guo, Y. Liu, Advances in flexible organic field-effect transistors and their applications for flexible electronics, Npj Flexible Electronics 2022 6:1. 6 (2022) 1–19. https://doi.org/10.1038/s41528-022-00133-3.

[23] X. Mu, W. Wang, C. Sun, D. Zhao, C. Ma, J. Zhu, M. Knez, Greatly increased electrical conductivity of PBTTT-C14 thin film via controllable single precursor vapor phase

infiltration, Nanotechnology. 34 (2022) 015709. https://doi.org/10.1088/1361-6528/AC96FA.

[24] V. Raja, Z. Hu, G. Chen, Progress in poly(2,5-bis(3-alkylthiophen-2-yl)thieno[3,2-b]thiophene) composites for thermoelectric application, Composites Communications. 27 (2021) 100886. https://doi.org/10.1016/J.COCO.2021.100886.

[25] Y. Huang, D.H. Lukito Tjhe, I.E. Jacobs, X. Jiao, Q. He, M. Statz, X. Ren, X. Huang, I. McCulloch, M. Heeney, C. McNeill, H. Sirringhaus, Design of experiment optimization of aligned polymer thermoelectrics doped by ion-exchange, Appl Phys Lett. 119 (2021) 111903. https://doi.org/10.1063/5.0055886.

[26] PDPP4T, PDQT | Polymer for Organic Solar Cells | Ossila, (n.d.). https://www.ossila.com/products/pdpp4t (accessed April 4, 2023).

[27] Y. Yang, Z. Liu, L. Chen, J. Yao, G. Lin, X. Zhang, G. Zhang, D. Zhang, Conjugated Semiconducting Polymer with Thymine Groups in the Side Chains: Charge Mobility Enhancement and Application for Selective Field-Effect Transistor Sensors toward CO and $H_2S$, Chemistry of Materials. 31 (2019) 1800–1807. https://doi.org/10.1021/ACS.CHEMMATER.9B00106/SUPPL_FILE/CM9B00106_SI_001.PDF.

[28] J. Ma, Z. Liu, Z. Wang, Y. Yang, G. Zhang, X. Zhang, D. Zhang, Charge mobility enhancement for diketopyrrolopyrrole-based conjugated polymers by partial replacement of branching alkyl chains with linear ones, Mater Chem Front. 1 (2017) 2547–2553. https://doi.org/10.1039/C7QM00307B.

[29] Vijila Chellappan, Adithya Kumar, S.A. khan, Pawan Kumar, Kedar Hippalgaonkar, Diagnosis of doped conjugated polymer films using hyperspectral imaging, Digital Discovery. 2 (2023) 471–480. https://doi.org/10.1039/D2DD00108J.

[30] A. Abutaha, V. Chellappan, P. Kumar, K. Hippalgaonkar, Linking Polaron Signatures to Charge Transport in Doped Thiophene Polymers, ACS Appl Energy Mater. 6 (2023) 3960–3969. https://doi.org/10.1021/ACSAEM.3C00149/ASSET/IMAGES/MEDIUM/AE3C00149_0006.GIF.

[31] D. Bash, Y. Cai, V. Chellappan, S.L. Wong, X. Yang, P. Kumar, J. Da Tan, A. Abutaha, J.J.W. Cheng, Y.F. Lim, S.I.P. Tian, Z. Ren, F. Mekki-Berrada, W.K. Wong, J. Xie, J. Kumar, S.A. Khan, Q. Li, T. Buonassisi, K. Hippalgaonkar, Multi-Fidelity High-Throughput Optimization of Electrical Conductivity in P3HT-CNT Composites, Adv Funct Mater. 31 (2021) 2102606. https://doi.org/10.1002/ADFM.202102606.

[32] R. Mamede, F. Pereira, J. Aires-de-Sousa, Machine learning prediction of UV–Vis spectra features of organic compounds related to photoreactive potential, Scientific Reports 2021 11:1. 11 (2021) 1–11. https://doi.org/10.1038/s41598-021-03070-9.

[33] Y. Xia, J.H. Cho, J. Lee, P.P. Ruden, C.D. Frisbie, Comparison of the mobility-carrier density relation in polymer and single-crystal organic transistors employing vacuum and liquid gate dielectrics, Advanced Materials. 21 (2009) 2174–2179. https://doi.org/10.1002/ADMA.200803437.

[34] C. Rudin, Why black box machine learning should be avoided for high-stakes decisions, in brief, Nature Reviews Methods Primers 2022 2:1. 2 (2022) 1–2. https://doi.org/10.1038/s43586-022-00172-0.

[35] W. Saeed, C. Omlin, Explainable AI (XAI): A systematic meta-survey of current challenges and future opportunities, Knowl Based Syst. 263 (2023) 110273. https://doi.org/10.1016/J.KNOSYS.2023.110273.

[36] H. Park, J.H. Lee, B-spline curve fitting based on adaptive curve refinement using dominant points, Computer-Aided Design. 39 (2007) 439–451. https://doi.org/10.1016/J.CAD.2006.12.006.

[37] K. Ahnert, M. Abel, Numerical differentiation of experimental data: local versus global methods, Comput Phys Commun. 177 (2007) 764–774. https://doi.org/10.1016/J.CPC.2007.03.009.

[38] H. Prautzsch, W. Boehm, M. Paluszny, Bézier and B-Spline Techniques, (2002). https://doi.org/10.1007/978-3-662-04919-8.

[39] D.F. Rogers, J.A. Adams, MATHEMATICAL ELEMENTS FOR COMPUTER Second Edition, New York. (1989).

[40] M.R.K. Mookiah, U. Rajendra Acharya, C.M. Lim, A. Petznick, J.S. Suri, Data mining technique for automated diagnosis of glaucoma using higher order spectra and wavelet energy features, Knowl Based Syst. 33 (2012) 73–82. https://doi.org/10.1016/J.KNOSYS.2012.02.010.

[41] M. Fey, J.E. Lenssen, F. Weichert, H. Muller, SplineCNN: Fast Geometric Deep Learning with Continuous B-Spline Kernels, 2018 IEEE/CVF Conference on Computer Vision and Pattern Recognition (CVPR). (2018) 869–877. https://doi.org/10.1109/CVPR.2018.00097.

[42] E.X. Tan, Y. Chen, Y.H. Lee, Y.X. Leong, S.X. Leong, C.V. Stanley, C.S. Pun, X.Y. Ling, Incorporating plasmonic featurization with machine learning to achieve accurate and bidirectional prediction of nanoparticle size and size distribution, Nanoscale Horiz. 7 (2022) 626–633. https://doi.org/10.1039/D2NH00146B.

[43] A.A. Guda, S.A. Guda, A. Martini, A.N. Kravtsova, A. Algasov, A. Bugaev, S.P. Kubrin, L. V. Guda, P. Šot, J.A. van Bokhoven, C. Copéret, A. V. Soldatov, Understanding X-ray absorption spectra by means of descriptors and machine learning algorithms, Npj Computational Materials 2021 7:1. 7 (2021) 1–13. https://doi.org/10.1038/s41524-021-00664-9.

[44] S.B. Torrisi, M.R. Carbone, B.A. Rohr, J.H. Montoya, Y. Ha, J. Yano, S.K. Suram, L. Hung, Random forest machine learning models for interpretable X-ray absorption near-edge structure spectrum-property relationships, Npj Computational Materials 2020 6:1. 6 (2020) 1–11. https://doi.org/10.1038/s41524-020-00376-6.

[45] G. Vilone, L. Longo, Notions of explainability and evaluation approaches for explainable artificial intelligence, Information Fusion. 76 (2021) 89–106. https://doi.org/10.1016/J.INFFUS.2021.05.009.

[46] P. Virtanen, R. Gommers, T.E. Oliphant, M. Haberland, T. Reddy, D. Cournapeau, E. Burovski, P. Peterson, W. Weckesser, J. Bright, S.J. van der Walt, M. Brett, J. Wilson, K.J. Millman, N. Mayorov, A.R.J. Nelson, E. Jones, R. Kern, E. Larson, C.J. Carey, İ. Polat, Y. Feng, E.W. Moore, J. VanderPlas, D. Laxalde, J. Perktold, R. Cimrman, I. Henriksen, E.A. Quintero, C.R. Harris, A.M. Archibald, A.H. Ribeiro, F. Pedregosa, P. van Mulbregt, A. Vijaykumar, A. Pietro Bardelli, A. Rothberg, A. Hilboll, A. Kloeckner, A. Scopatz, A. Lee, A. Rokem, C.N. Woods, C. Fulton, C. Masson, C. Häggström, C. Fitzgerald, D.A. Nicholson, D.R. Hagen, D. V. Pasechnik, E. Olivetti, E. Martin, E. Wieser, F. Silva, F. Lenders, F. Wilhelm, G. Young, G.A. Price, G.L. Ingold, G.E. Allen, G.R. Lee, H. Audren, I. Probst, J.P. Dietrich, J. Silterra, J.T. Webber, J. Slavič, J.

Nothman, J. Buchner, J. Kulick, J.L. Schönberger, J.V. de Miranda Cardoso, J. Reimer, J. Harrington, J.L.C. Rodríguez, J. Nunez-Iglesias, J. Kuczynski, K. Tritz, M. Thoma, M. Newville, M. Kümmerer, M. Bolingbroke, M. Tartre, M. Pak, N.J. Smith, N. Nowaczyk, N. Shebanov, O. Pavlyk, P.A. Brodtkorb, P. Lee, R.T. McGibbon, R. Feldbauer, S. Lewis, S. Tygier, S. Sievert, S. Vigna, S. Peterson, S. More, T. Pudlik, T. Oshima, T.J. Pingel, T.P. Robitaille, T. Spura, T.R. Jones, T. Cera, T. Leslie, T. Zito, T. Krauss, U. Upadhyay, Y.O. Halchenko, Y. Vázquez-Baeza, SciPy 1.0: fundamental algorithms for scientific computing in Python, Nature Methods 2020 17:3. 17 (2020) 261–272. https://doi.org/10.1038/s41592-019-0686-2.

[47] C. de Boor, A Practical Guide to Splines - Revised Edition, Springer, 2001. https://link.springer.com/book/9780387953663 (accessed June 22, 2023).

[48] F. Pedregosa FABIANPEDREGOSA, V. Michel, O. Grisel OLIVIERGRISEL, M. Blondel, P. Prettenhofer, R. Weiss, J. Vanderplas, D. Cournapeau, F. Pedregosa, G. Varoquaux, A. Gramfort, B. Thirion, O. Grisel, V. Dubourg, A. Passos, M. Brucher, M. Perrot andÉdouardand, andÉdouard Duchesnay, Fré. Duchesnay EDOUARDDUCHESNAY, Scikit-learn: Machine Learning in Python, Journal of Machine Learning Research. 12 (2011) 2825–2830. http://jmlr.org/papers/v12/pedregosa11a.html (accessed April 5, 2023).

[49] A. Baratloo, M. Hosseini, A. Negida, G. El Ashal, Part 1: Simple Definition and Calculation of Accuracy, Sensitivity and Specificity, Emerg (Tehran). 3 (2015) 48–9. https://pubmed.ncbi.nlm.nih.gov/26495380/ (accessed April 5, 2023).